\documentclass[preprint]{aastex}
\usepackage{natbib}
\shorttitle{Mid-IR Study of MWC 1080 and HD 259431}
\shortauthors{Li et al.}

\begin{document}

\title{The Immediate Environments of Two Herbig Be Stars: MWC 1080 and HD 259431}

\author{Dan Li, Naib\'{i} Mari\~{n}as, and Charles M. Telesco}
\affil{Department of Astronomy, University of Florida,
    Gainesville, FL 32611, USA}
\email{d.li@ufl.edu}

\begin{abstract}
Deep mid-infrared (10-20 $\micron$) images with sub-arcsec resolution were obtained for two Herbig Be stars, MWC 1080 and HD 259431, to probe their immediate environments. Our goal is to understand the origin of the diffuse nebulosities observed around these two very young objects. By analyzing our new mid-IR images and comparing them to published data at other wavelengths, we demonstrate that the well extended emission around MWC 1080 traces neither a disk nor an envelope, but rather the surfaces of a cavity created by the outflow from MWC 1080A, the primary star of the MWC 1080 system. In the N-band images, the filamentary nebulosities trace the hourglass-shaped gas cavity wall out to $\sim$0.15 pc. This scenario reconciles the properties of the MWC 1080 system revealed by a wide range of observations. Our finding confirms that the environment around MWC 1080, where a small cluster is forming, is strongly affected by the outflow from the central Herbig Be star. Similarities observed between the two subjects of this study suggest that the filamentary emission around HD 259431 may arise from a similar outflow cavity structure, too.

\end{abstract}


\section{Introduction}
Herbig Ae/Be stars are intermediate-mass pre-main-sequence stars \citep{herbig1960, waters1998}. Their spectral energy distributions (SEDs) exhibit infrared excesses due to emission from dust grains present in circumstellar disks and/or envelopes. Disks around Herbig Ae/Be stars have been spatially resolved at many wavelengths \citep[e.g.,][]{grady2005, doucet2006, isella2007, hashimoto2011}, and there is a growing understanding of them based on the comparison of the observed SED and morphologies with theoretical models that include a variety of disk architectural elements such as flaring \citep{marinas2006, doucet2007, okamoto2009} and disk holes or gaps \citep[e.g.,][]{monnier2005, chen2012}. However, observations also suggest that not all of the infrared excess detected around Herbig stars is confined to a disk, per se, or a simple contiguous envelope, but rather originates in a more complex circumstellar environment \citep[e.g.,][]{verhoeff2012}.

In particular, we draw attention to two very young ($<$1.0 Myr) Herbig Be stars, MWC 1080 and HD 259431 \citep{levreault1988, bouret2003}. In a previous study, \citet[][P02 hereafter]{polomski2002} observed these objects in the mid-infrared 10- and 20-$\micron$ bands, and revealed complex filamentary structures around them. The origin of the extended features remains unclear though. The morphologies of the extended emission mimic spiral arms, which are suggestive of tidal interactions commonly seen in massive disks \citep[e.g.,][]{hashimoto2011, tang2012, grady2013}. However, given the distance to these objects, the linear scale of the observed structures is too big to be associated with the general picture of circumstellar disks. In order to understand the nature of the extended emission, which may provide insight into the relationship between Herbig Be stars and their environments, we obtained very deep images of MWC 1080 and HD 259431 in the mid-IR with sub-arcsec spatial resolution. We aim to propose a scenario that is reconciled with the complex structures revealed by our observations as well as others acquired at shorter or longer wavelengths for these two objects.

This paper is organized as follows: in Section 2, we briefly review the two objects of this study. Section 3 describes the observations and data reduction procedures. We analyze the data in Section 4. In Section 5, we discuss our results and compare them with other observations. Section 6 contains the main conclusions.

\section{Sources}
\subsection{MWC 1080}
MWC 1080 (=V628 Cas) was classified as a Herbig Be star by \citet{cohen1979} (see also \citealt{hillenbrand1992}). High resolution observations revealed that it is a triple system containing a primary (MWC 1080A) that itself is an eclipsing binary with a period of 2.9 days \citep{shevchenko1994}, with a close companion (MWC 1080B) $\sim$0\farcs76 west of the primary \citep{leinert1997, polomski2002}. MWC 1080B is also a Herbig Ae/Be star of $\sim250\,L_{\odot}$ \citep{leinert1997}. Confusingly, another object found $\sim$5\farcs2 east of the primary is sometimes regarded as the third companion, and referred to as MWC 1080E (P02). MWC 1080 is associated with a small cluster embedded in the dark cloud LDN 1238 at a distance of 2.2 kpc \citep{abraham2000}. More than 40 young, low-mass stars and 32 dense gas clumps have been detected within 0.3 pc of MWC 1080A \citep[][hereafter W08]{wang2008}. Single-dish $\mathrm{^{13}CO}$, $\mathrm{C^{18}O}$, and $\mathrm{CS}$ observations show that the MWC 1080 system is associated with more than 1000 $M_{\odot}$ of molecular material within 1 pc \citep{hillenbrand1995}, and 10 $M_{\odot}$ within 0.08 pc \citep{fuente2002}. 

As the most massive resident in its neighborhood, MWC 1080 exerts considerable influence on the environment where it forms, serving as a rare example of small star forming regions dominated by intermediate mass B stars (in contrast to more destructive O stars or OB associations). Clear evidence of the influence asserted by MWC 1080 can be seen in the $\mathrm{CS}\,(J=2\rightarrow 1)$, $\mathrm{^{13}CO}\,(J=1\rightarrow 0)$, and $\mathrm{C^{18}O}\,(J=1\rightarrow 0)$ maps that trace the gas distribution (see Figure 2 in W08). The data reveal a single-lobe cavity of size 0.3 pc $\times$ 0.05 pc along PA = 45\arcdeg\, thought to be cleared by the bipolar outflow from MWC 1080. Almost all of the young, low-mass cluster stars are within this cavity, or close to its edge, implying that they are not background or foreground objects but members of the MWC 1080 cluster, possibly only emerging from the cloud after it was dispersed by the MWC 1080A's strong bipolar outflow.

\subsection{HD 259431}
HD 259431 (=MWC 147, V700 Mon) is a Herbig Be star associated with the reflection nebula NGC 2247 at a distance of 290 pc \citep{vanleeuwen2007}. P02 observed this object in the mid-IR and constructed the SED. The overall shape and the 3 - 10 $\micron$ spectral index are consistent with the model of a moderately flared circumstellar disk. However, there is no direct evidence (i.e., resolved images) for the disk, and the possibility of a significant contribution to the IR excess from a circumstellar envelope or halo cannot be ruled out. For example, analyses with the $FUSE$ far ultraviolet spectra (905-1187 \AA) indicate that the material observed is most likely related to a large circumstellar envelope, which is presumably the remnant of the parent molecular clouds \citep{bouret2003, martin-zaidi2008}.

\section{Observation and data reduction}
We observed MWC 1080 and HD 259431 in 2005 using Michelle, the facility mid-IR camera at Gemini North. The total field of view was 32\farcs16 $\times$ 24\farcs12 with a pixel scale of 0\farcs1005. Two N-band filters, Si-5 (11.6 $\micron$, $\Delta\lambda=$ 0.9 $\micron$) and N' (11.2 $\micron$, $\Delta\lambda=$ 2.1 $\micron$), and one Q-band filter, Qa (18.5 $\micron$, $\Delta\lambda=$ 0.9 $\micron$) were used. \citet{acke2010} found strong PAH features at 8.6 and 11.3 $\mu m$ in the mid-IR spectrum of HD 259431. In order to determine the N-band continuum flux more accurately, we took an additional image of HD 259431 in 2011 using Michelle with the Si-4 filter (10.3 $\micron$, $\Delta\lambda=$ 1.0 $\micron$). 

Several mid-IR standard stars (i.e., the Cohen stars, \citealt{cohen1999}) of spectral type K0 to K2 were observed for flux calibration and monitoring of the point spread function (PSF) (Table. \ref{tab:obs_log}). The angular separations between the program stars and the standards were less than 13\arcdeg\, in all cases. Standard chop-and-nod procedures were applied to correct for the thermal background from the telescope and the atmosphere. The chop/nod throw was 15\arcsec, the maximum throw allowed by the Gemini North telescope. 

Raw data in FITS format were reduced with \emph{iDealCam} \citep{li2013} following standard routines (chop correction, bad frame removal, frame registration, etc.). In stacked images, non-uniform background and cross-talk patterns are apparent, and custom treatments were used to correct for these image artifacts. Specifically, the background residual is measured in each channel and subtracted. The cross-talk pattern is modeled using the rightmost and leftmost channels, where fluxes from the source are negligible, and then subtracted from each channel. The resultant images are presented in Fig.  \ref{fig:contours}.

Several point sources are found in the close vicinity of MWC 1080, most of which have near-IR counterparts in the K'-band image presented in W08. The brightest one of them is MWC 1080E at 0\farcs51 arcsec east of MWC 1080. There is also a point source (``Source B'') near HD 25943, located at 3\farcs1 arcsec NNW. Flux densities of all point sources that have been detected with statistical significance greater than 3 $\sigma$ are measured using aperture photometry and summarized in Table \ref{tab:hd259431_photometry} and \ref{tab:mwc1080_photometry}.

\section{Results}

\subsection{Mid-infrared morphology}\label{sec:mid_ir_morphology}
\begin{figure}
\plotone{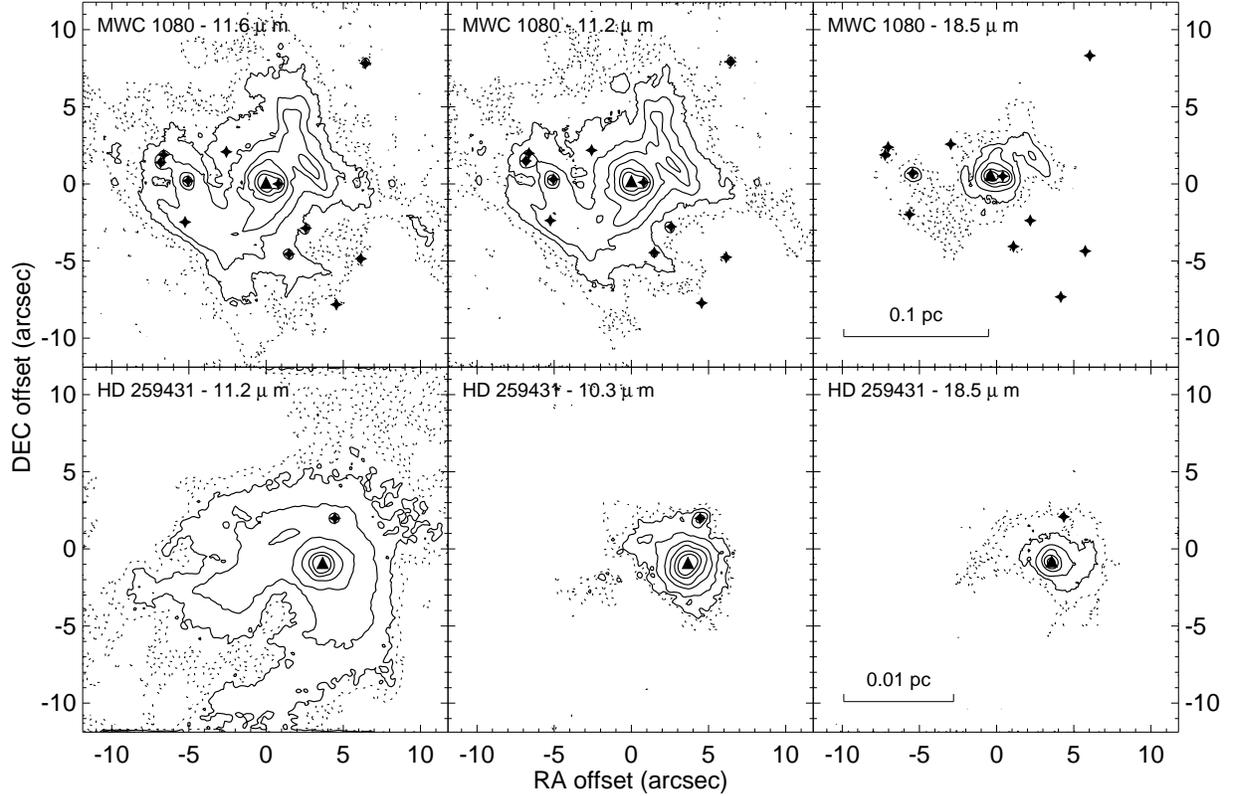}
\caption{Upper panels: images (contours) of MWC 1080 at 11.6, 11.2, and 18.5 $\micron$. The position of MWC 1080A is indicated by a solid triangle. Solid stars are point sources, including MWC 1080 B and E, visually identified in the 11.6 $\micron$ image. The lowest contours (thin line) are 1 $\sigma$ (3.7, 3.3, 38.3 mJy/arcsec$^2$ at three wavelengths respectively). Lower panels: images (contours) of HD 259431 at 11.2, 10.3, and 18.5 $\micron$.  The lowest contours (thin line) are 1 $\sigma$ (1.6, 2.8, 27.4 mJy/arcsec$^2$ at three wavelengths respectively). The position of HD 259431 is labeled by a solid triangle. In all images, solid contours are 3$\times$, 9$\times$, 27$\times$, 81$\times$, 243$\times$, and 741$\times\sigma$. \label{fig:contours}}
\end{figure}

\begin{figure}
\plotone{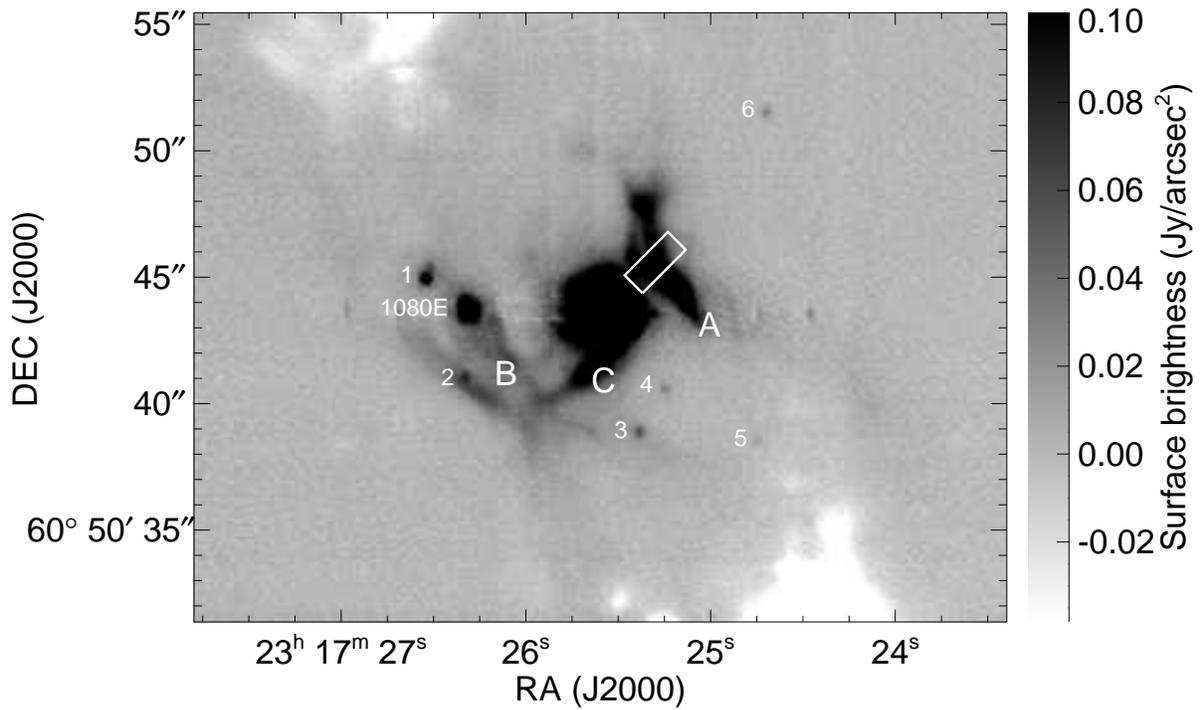}
\caption{N'-band (11.2 $\micron$ images of MWC 1080 in reversed grayscale. Point sources are labeled by their names or numbers (Source 1-6). The three regions discussed in Section \ref{sec:mid_ir_morphology} are labeled by A (``NW wall'' ), B (``SE filaments''), and C (``south finger''). The area enclosed by the white rectangle is used for the analysis of color temperature and optical depth in Section \ref{sec:t_tau}.\label{fig:mwc1080_n}}
\end{figure}

The mid-IR morphology of the MWC 1080 system is very complex. Filamentary emission, presumably originating in warm dust grains and/or PAH (polycyclic aromatic hydrocarbons) molecules, extends from MWC 1080 to about 15\arcsec\, ($\sim$0.15 pc) at 11.2 and 11.6 $\micron$, or 7\arcsec\, ($\sim$0.07 pc) at 18.5 $\micron$. In contrast to previous observations that reveal the dominant emission to be extended along the NW-SE direction (e.g., P02), our newer, deeper images show considerable lower-level emission extending along the NE-SW direction (see the 1 $\sigma$ contours in Fig. \ref{fig:contours}). MWC 1080A and B are the brightest mid-IR sources in the field of view, contributing approximately 80\% of the total flux. Another conspicuous structure is a ``wall'' that lies about 0.03 pc NW of MWC 1080 (referred to as the ``NW wall'' hereafter, Fig. \ref{fig:mwc1080_n}), extending for $\sim$0.1 pc from NE to SW. This wall is clearly detected in both N and Q bands. SE of MWC 1080, there are emitting filaments (``SE filaments'') that are less luminous but more extended compared to the NW wall. The SE filaments are apparent at 11 $\micron$, but barely detected at 18.5 $\micron$. In N-band images, there is a finger-like structure (``south finger'') that emerges from MWC 1080A/B, pointing to the southeast and eventually reaching the SE filaments.  Emission from this feature is also visible at 18.5 $\micron$. The SE filaments and south finger together form a prominent structure that, to a degree, mimics a spiral arm in the N-band images.

MWC 1080, the NW wall and SE filaments together merge into the dominant emission complex extending for $\sim$0.1 pc along PA = -50\arcdeg\, (see the 3 $\sigma$ contours in Fig. \ref{fig:contours}). This region is relatively bright and has been imaged before (P02). \citet{fuente2003} observed a similar structure in their dust continuum maps at 1.4 and 2.7 mm, although the origin of it was unclear at that time. As discussed below, our images suggest that the elongation does not trace a simple ``disk'' or ``envelope'' surrounding the MWC 1080 system.

HD 259431 also shows a complex morphology at the wavelengths we observed. Thanks to the better sensitivity, our new Michelle image reveals more extended structures than do previous observations (P02). In the 11.2 $\micron$ image, two filamentary structures are noticeable, extending SE for $\sim$ 15'' (equivalent to 4350 AU, or 0.021 pc, at 290 pc) from the brightest central source. These two arm-like filaments merge to form part of a bow-shaped structure centered on HD 259431. The overall appearance of the HD 259431 system is very asymmetric, with the emission being heavily skewed to the SE. Much of the extended emission may be from PAH particles, since it is much less evident at 10.3 $\micron$, a filter that was chosen to avoid the 11.3 $\micron$ PAH band. This is also consistent with previous spectroscopic studies of this object \citep{wooden1994, acke2010}.

Whether the Source B, which is observed at 3\farcs1 away from HD 259431 along P.A. $\sim$ -14\fdg7, is a foreground or background star or a close companion of HD 259431 is unclear. The 18-$
\micron$/10-$\micron$ flux density ratio is consistent with a 1000 K blackbody, suggesting that it is also a young object like the primary star. However, the proper motion of HD 259431 spanning several years is not sufficient to rule out the other possibility. 

\subsection{Source size}
MWC 1080A and its companion MWC 1080B are clearly resolved as two objects in the Michelle images. The angular size of MWC 1080A is measured using quadratic subtraction as outlined in \citet{marinas2006, marinas2011}. The result indicates that MWC 1080A itself is not spatially resolved (i.e., the source size measured at the FWHM brightness contour level is not greater than three times the error) at any wavelength we observed. The 3 $\sigma$ upper limits of the angular size are 57, 42, and 48 mas at 11.2, 11.6, and 18.5 $\micron$, respectively, corresponding to disk diameters of 125, 92, and 106 AU at the distance of 2.2 kpc. Using the continuum map at 1.4 mm, \citet{fuente2003} concluded that the unresolved MWC 1080 disk is less than $\sim$2000 AU in diameter. Our results improve this constraint considerably, although the mid-IR emission traces the small grains (sub-micron- to micron-sized) on the disk surface, while mm observations are sensitive to bigger, cooler grains closer to the disk mid-plane. The upper limit of the disk size derived here is in good agreement with $r_{out}\sim77$ AU, a value implied by multiple wavelength photometry and SED modeling of MWC 1080A \citep{alonso-albi2009}.

The disk around HD 259431 is not spatially resolved either. Derived 3-$\sigma$ upper limits of the angular size are 87, 81, and 57 mas at 11.2, 11.6, and 18.5 $\micron$, respectively, corresponding to disk diameters of 25, 23, and 17 AU at the distance of 290 pc. \citet{verhoeff2012} studied a sample of Herbig Be stars with mid-IR imaging/spectroscopy and SED modeling, and concluded that Herbig Be stars have flatter disks than the Herbig Ae stars and that they are likely truncated on the outside by photoevaporation. If that is the case, then it is not surprising that neither disk in our sample is spatially resolved. 

\subsection{Color temperature and optical depth}\label{sec:t_tau}
For optically thin emission, the observed flux density is proportional to the product of dust temperature $T_d$ and (emissive) optical depth $\tau$ (ignoring foreground extinction). For an observation at wavelength $\lambda$, we may write:

\begin{equation}
F_{\lambda}=\Omega \cdot \tau_{\lambda} \cdot B_{\lambda}(T_d),
\end{equation}
where $F_{\lambda}$ is the observed flux density, $\Omega$ is the solid angle corresponding to one Michelle pixel, $\tau_{\lambda}$ is the optical depth at wavelength $\lambda$, and $B_{\lambda}(T_d)$ is the Planck function. With two continuum observations in N and Q bands, we are able to solve for $T_d$ and $\tau$ iteratively. 

For MWC 1080, we use 11.2 and 18.5 $\micron$ images to calculate the 11.2/18.5 color temperature and the optical depth. For HD 259431, the spectrum of which shows strong PAH feature at 11.3 $\micron$, we use 10.3 and 18.5 $\micron$ images instead. The values of $A_v$ given in Table \ref{tab:objects} are adopted for extinction correction, and we assume $A_{11.2}/A_v=0.054$, $A_{10.3}/A_v=0.078$ \citep{rieke1985extinctionlaw}, and $\tau\propto1/\lambda$ \citep{williams2011}. We also correct for the Si emission feature that has been found for MWC 1080 \citep{sakon2006}. For simplicity, we assume a uniform flux density ratio (0.25) between the Si emission and underlying continuum across the entire field of view. This ratio is estimated from the N band spectra of MWC 1080A/B and the NW wall presented in \citet{sakon2006}. To compensate for the difference in spatial resolution between 11.2 and 18.5 $\micron$, each image is smoothed with a Gaussian core with FWHM equal to the spatial resolution measured at the other wavelength.

\begin{figure}
\plottwo{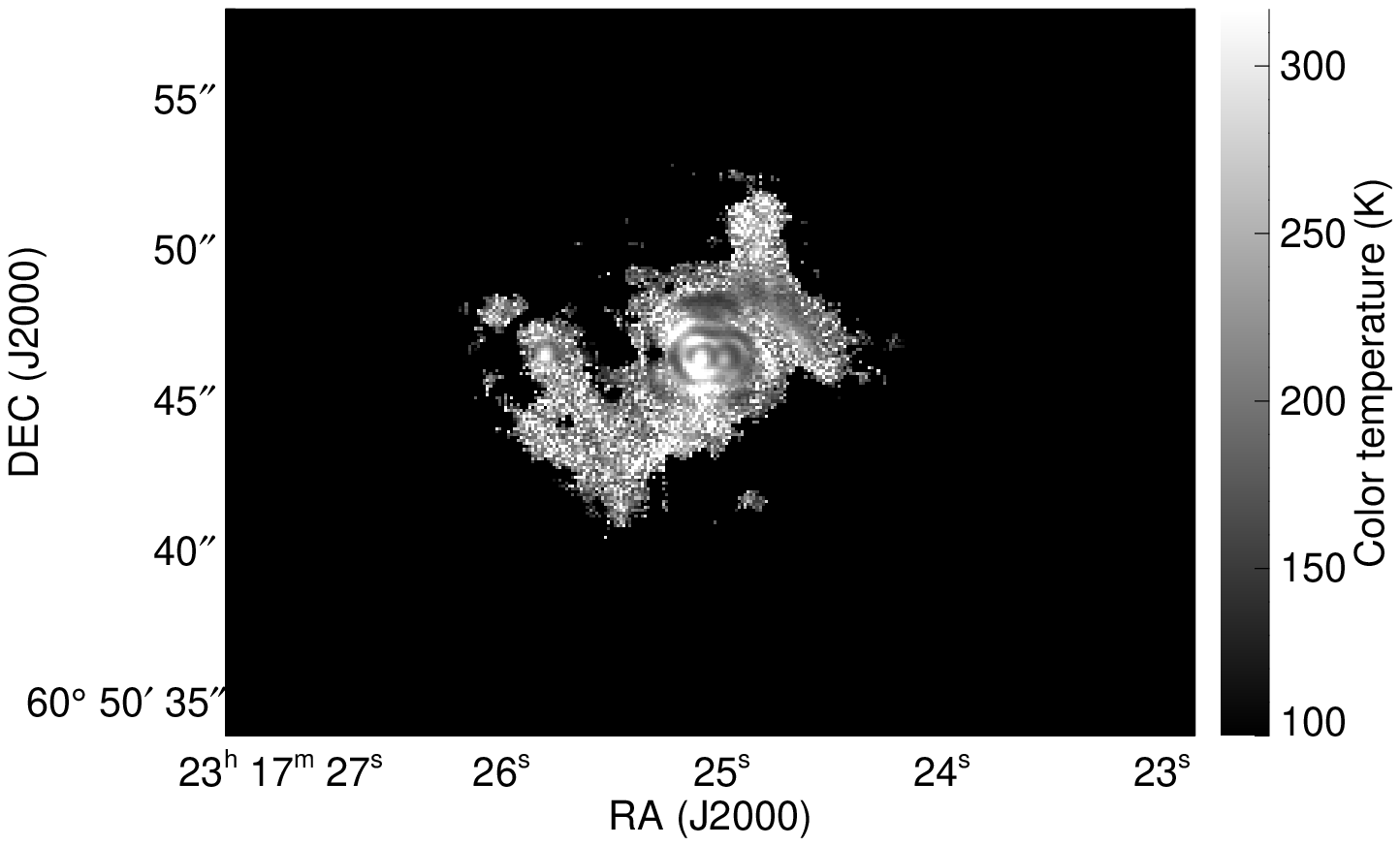}{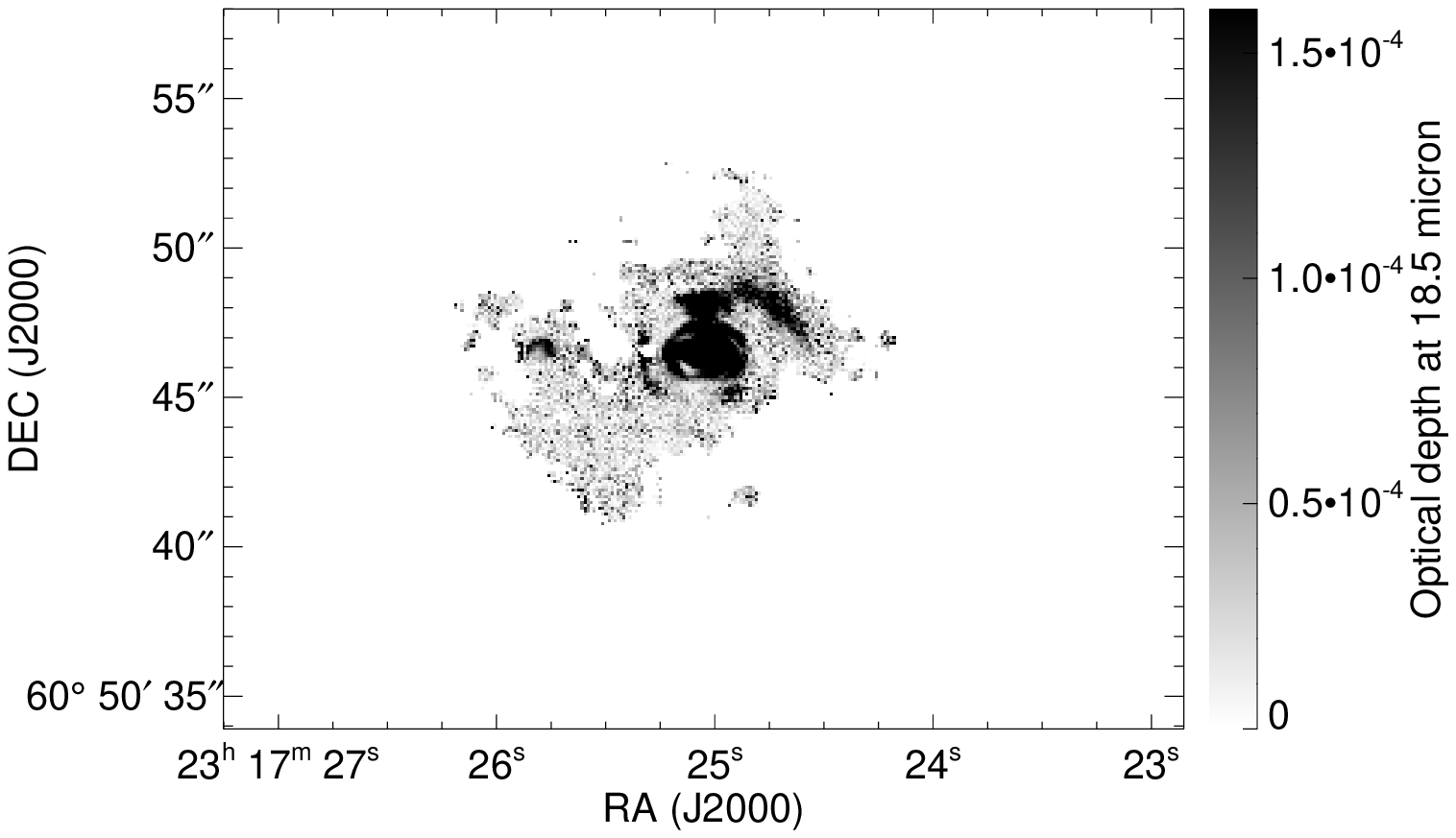}
\caption{Maps of color temperature (left) and optical depth ($\tau_{18.5}$) computed with the 11.2 $\micron$ and 18.5 $\micron$ images of MWC 1080.\label{fig:mwc1080_t_and_tau_maps}}
\end{figure}

The maps of the color temperature and emitting optical depth ($\tau_{18.5}$) are presented in Fig. \ref{fig:mwc1080_t_and_tau_maps}. The area where $T_d$ and $\tau_{18.5}$ can be computed is roughly defined by the 1 $\sigma$ contour in the 18.5 $\micron$ image. We first check $T_d$ and $\tau_{18.5}$ around MWC 1080A \& B, where the signal-to-noise ratio is high. We find hotter material ($\sim$360 K) around MWC 1080A and cooler material ($\sim$260 K) around MWC 1080B. MWC 1080A also appears to be slightly less embedded than B ($\tau_{1080A}/\tau_{1080B}\sim0.8$). These results are consistent with those of P02.

\begin{figure}
\plotone{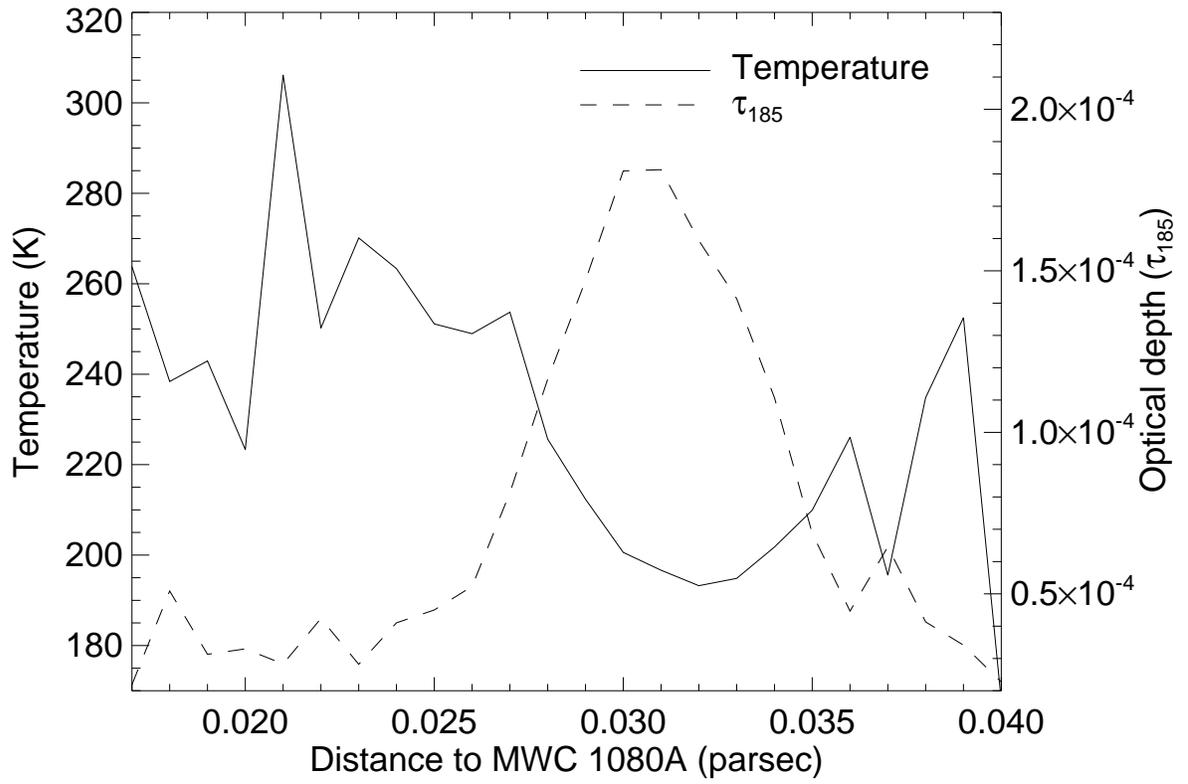}
\caption{Mid-IR color temperature and optical depth as functions of the distance to MWC 1080A. Data are measured in the ``NW wall'' region highlighted in Fig. \ref{fig:mwc1080_n}.\label{fig:mwc1080_t_and_tau}}
\end{figure}

We emphasize that the optical depth and temperature values presented in the maps are line-of-sight average values. They are inaccurate to some degree, because both quantities are weighted by the grain properties, which vary across the field of view. Nevertheless, these maps should permit an examination of the region's key features. The ring-like structures around MWC 1080, visible in both $T_d$ and $\tau_{18.5}$ maps, are artifacts associated with the diffraction rings. Excluding them, the most prominent feature in the $\tau_{18.5}$ map is a high-optical-depth region coincident with the NW wall. In the same part of the $T_d$ map, the area facing MWC 1080 is significantly hotter than that behind it. We consider it evidence for the NW wall region being heated externally by MWC 1080. To illustrate this point further, a rectangular region extending perpendicular to the NW wall is highlighted in Fig. \ref{fig:mwc1080_n}, with profiles of the temperature and optical depth along the region's long axis plotted in Fig. \ref{fig:mwc1080_t_and_tau}. The profiles describe how the dust temperature and optical depth vary as functions of the projected distance to MWC 1080A. Notably, the peak of the dust temperature distribution is not coincident with that of the optical depth. The highest temperature occurs closer to MWC 1080, while the peak optical depth is found at a place farther away. Such a correlation between the temperature and optical depth is expected for dense clouds being heated externally, and has been observed in ionization fronts facing luminous OB stars \citep[e.g.,][]{telesco1996}. Based on this analysis, we conclude that there is no embedded sources inside the NW wall region. 

Although the the $T_d$ map across the SE filaments is very noisy, no apparent temperature gradient is found, and the averaged temperature in this region is $\sim$200 K. To compare the color temperature derived from observations and that calculated theoretically, we use the energy balance equation (assuming spherical grains):

\begin{equation}\label{eq:energy_balance}
\int Q_{\nu }^{\textup{abs}}\left ( a \right )\cdot J_{\nu }d\nu =\int Q_{\nu }^{\textup{abs}}\left ( a \right )\cdot B_{\nu }\left ( T_{d} \right )d\nu \, , 
\end{equation}
where $J_{\nu }$ is the average of the radiation intensity over all directions, and $Q_{\nu }^{\mathrm{abs}}$ is the absorption efficiency. At infrared wavelengths, we have a good approximation for the integral on the right (i.e., the emission term), $Q_{\nu }^{\textup{abs}}\simeq 10^{-23}a\nu^2$, with $a$ being the grain size in cm \citep{kruegel2003book}. For the absorption term on the left, one may put $Q_{\nu }^{\mathrm{abs}}\simeq 1$, which is valid for early-type stars like MWC 1080. For $J_{\nu}$, we assume it to be a diluted blackbody radiation. Eventually, one can get

\begin{equation}
T_d(\textup{K})=288\cdot \left ( \frac{L}{L_{\odot }} \right )^{\frac{1}{6}}\cdot \left ( \frac{a}{1\, \mu m} \right )^{-\frac{1}{6}}\cdot \left ( \frac{r}{1\, \textup{AU}} \right )^{-\frac{1}{3}}\, .
\end{equation}
Assuming $L=10^4L_{\odot}$, $a=0.1\,\mathrm{\mu m}$, and $r=10,000$ AU, we get $T_d\simeq100$ K. This is lower than the values we got from observations by a factor $\sim$2. The difference implies either very small grains ($<$ 0.1 $\micron$) presenting in the emitting regions, or additional sources embedded in SE filaments heating the environment but not detected by us. Indeed, there are several near-IR-identified young stars found in the SE filaments that are likely to be able to account for it (see Figure 1 of W08). We note that, however, the dust temperature calculated here is based on a strong assumption of optically thin emission, which may not be valid everywhere around MWC 1080.

\section{Discussion}
\subsection{MWC 1080}
\begin{figure}
\plotone{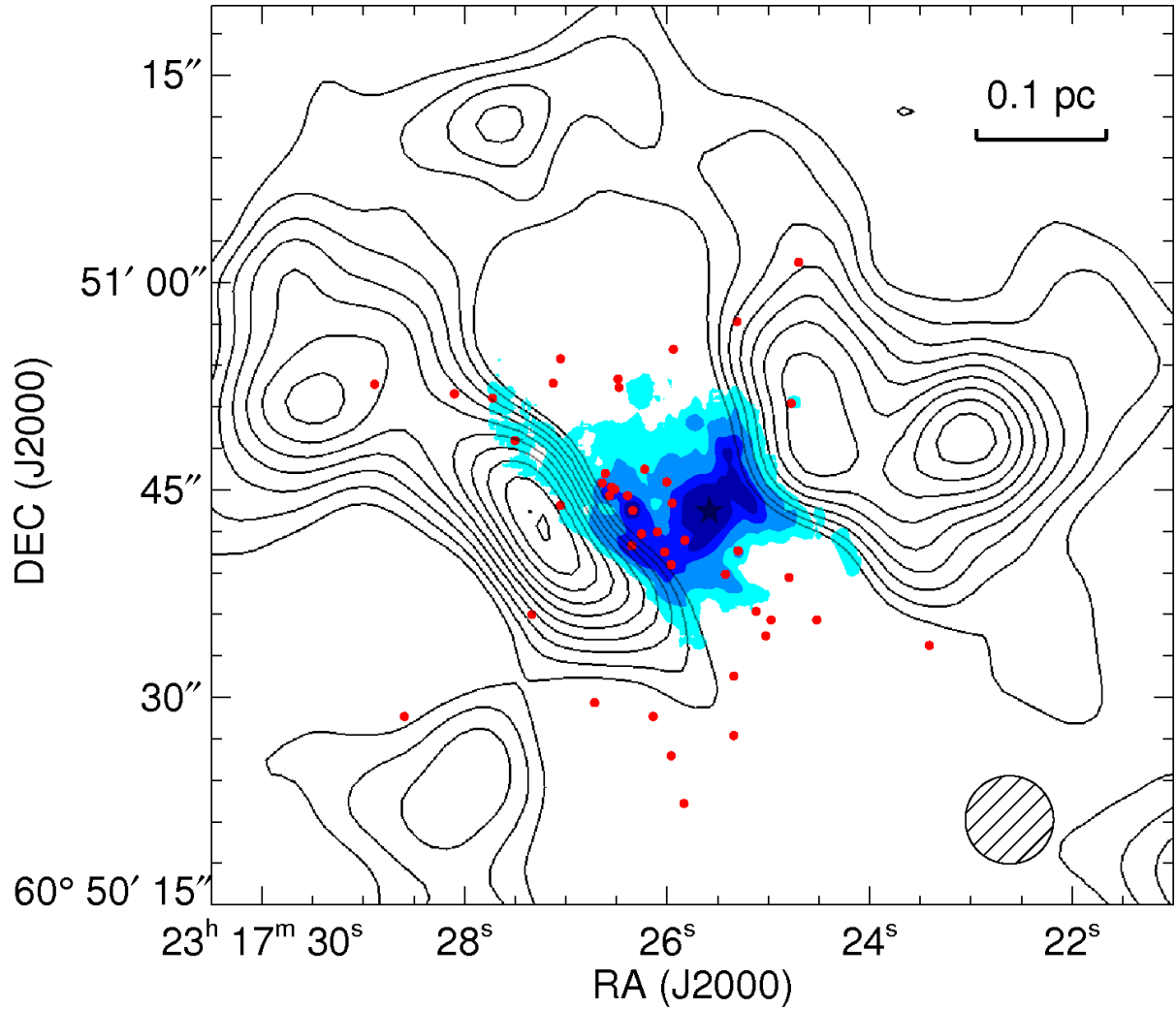}
\caption{The $^{13}$CO contours (black) of MWC 1080 overlaid with Gaussian-smoothed 11.2 $\mathrm{\mu m}$ image (filled contours). The CO contours are reproduced from \citet{wang2008} by permission of the AAS. The beam size of the CO maps is 6.4\arcsec $\times$ 6.3\arcsec. Filled points indicate the near-IR-identified cluster members \citep{wang2008}. See the electronic edition of the Journal for a color version of this figure.\label{fig:mwc1080_on_13co}}
\end{figure}

The stellar environment that gives rise to the mid-IR morphology of the MWC 1080 system can be better understood by comparing the distributions of dense gas and warm dust in the vicinity of MWC 1080. In Fig. \ref{fig:mwc1080_on_13co}, we show the $\mathrm{^{13}CO}\,J=1\rightarrow 0$ ($\nu=$110.201 GHz) map of W08 made at the Berkeley Illinois Maryland Association (BIMA) array with a 6.4\arcsec $\times$ 6.3\arcsec beam. In the $\mathrm{^{13}CO}$ map, there is a biconical cavity around MWC 1080, a signature of bipolar outflows. The large gas loop to the north is the boundary of the northern outflow, with only part of the southern loop nearest the star MWC 1080 being evident. Overlaid onto the $\mathrm{^{13}CO}$ map are the Michelle 11.2 $\micron$ contours. The mid-IR morphology of the MWC 1080 system is well correlated with the cavity structure. Near the NW wall and the SE filaments, our mid-IR contours align precisely with the $\mathrm{^{13}CO}$ velocity-integrated contours, implying that the observed mid-IR fluxes are emitted by dust located close to the outflow cavity wall, where the material is externally heated by the stellar radiation. Diffuse mid-IR emission is not confined to the NW wall and the SE filaments, but is extended within the cavity, probably originating from surfaces of the gas clouds that are behind the cavity along the line-of-sight. For the NW wall and the SE filaments, however, our line-of-sight runs roughly parallel to the emitting surfaces, thus encountering more optical depth. This effect accounts for the higher mid-IR surface brightness of the NW wall and the SE filaments compared to other regions around MWC 1080.
 
\begin{figure}
\plotone{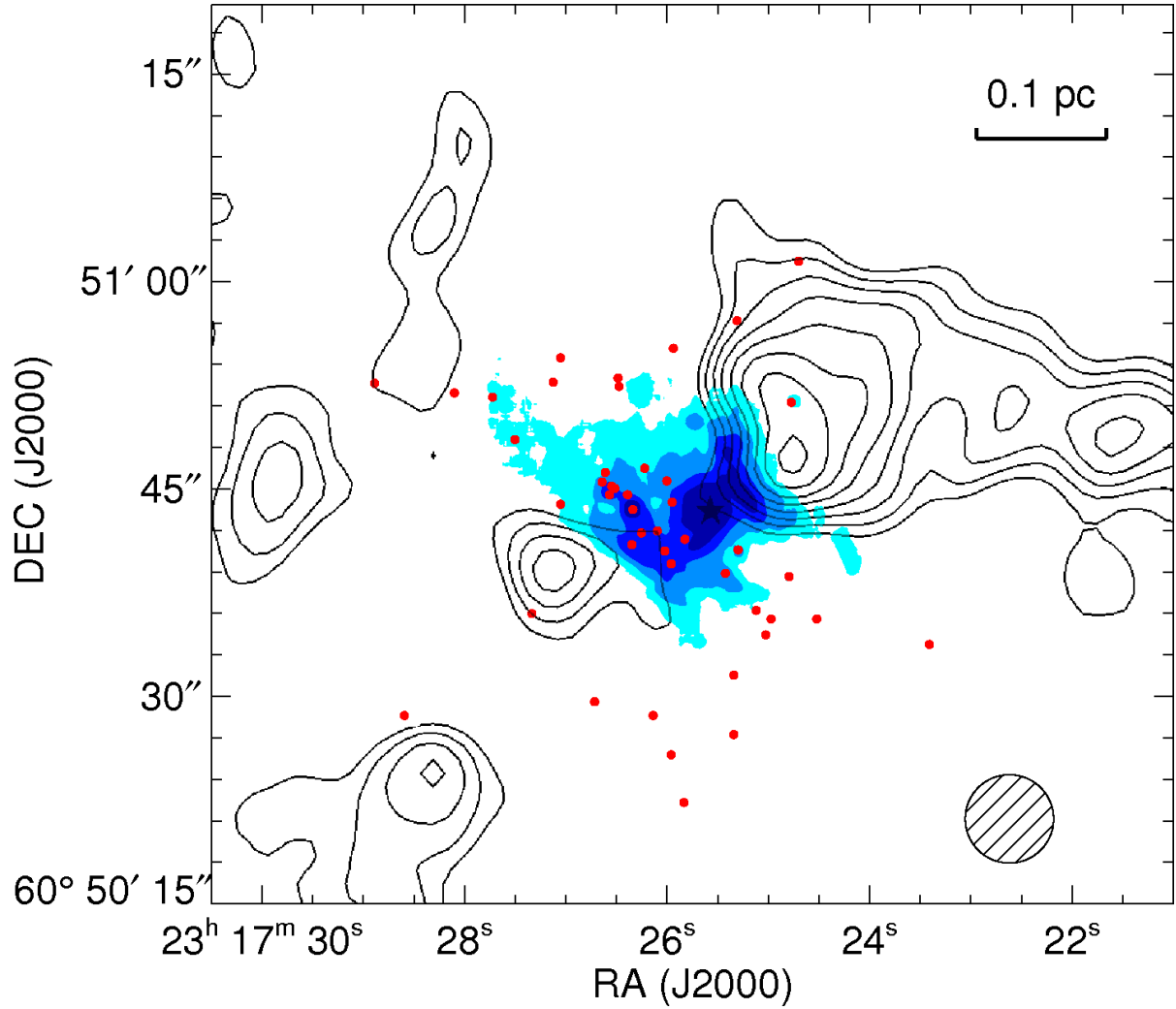}
\caption{Same as Fig. \ref{fig:mwc1080_on_13co} but overlaid with C$^{18}$O contours (black), which are reproduced from \citet{wang2008} by permission of the AAS. See the electronic edition of the Journal for a color version of this figure.\label{fig:mwc1080_on_c18o}}
\end{figure}

In Fig. \ref{fig:mwc1080_on_c18o}, the $\mathrm{C^{18}O}\,J=1\rightarrow 0$ ($\nu=$109.782 GHz) map from W08 is also plotted overlaid with our 11.2. $\micron$ contours. Once again, the mid-IR morphology is very complementary with the gas distribution. Judging from the $\mathrm{C^{18}O}$ map, the gas dispersal around MWC 1080 is not symmetric: the gas lobe to the NW appears to be closer in projection to MWC 1080 than the SE lobe. Therefore, assuming the relative projected distances reflect the actual distances, the mixture of gas and dust on the NW side of MWC 1080 undergoes more intense heating by MWC 1080, thus accounting for its stronger emission. This is in agreement with what we see in the mid-IR images: the NW wall is significantly brighter than the SE filaments.

The scenario pictured above can also explain why the spatial distribution of mid-IR point sources around MWC 1080 is highly nonuniform: all of them except one (Source 6) are found within an elongated area along the NE-SW axis of MWC 1080 system. A similar distribution is found in the near-IR K' image (Fig. \ref{fig:contours}). Figure \ref{fig:mwc1080_on_13co} and \ref{fig:mwc1080_on_c18o} suggest that the extinction toward this area is low compared to other regions around MWC 1080, thus leaving the cluster stars visible in the mid-IR. This is similar to the case of Orion Nebula, but in a smaller scale.

The $\mathrm{CS}$ observations made by W08 provide quantitative evidence that the intrinsic distribution of MWC 1080 cluster members may not be that nonuniform, as many stars may be hidden behind the natal material surrounding the cluster. The relatively unoccupied region to the SE and NW of MWC 1080 corresponds to two gas clumps, the column densities ($N_{H}$) of which are 33.9$\times$10$^{22}$ cm$^{-2}$ and 7.88-12.79$\times$10$^{22}$ cm$^{-2}$, respectively (W08). Assuming the interstellar extinction law of \citet{rieke1985extinctionlaw} and $A_{V}/N_{H}\approx5.35\times10^{-22}$ mag cm$^2$ for $R_{V}=3.1$ \citep{bohlin1978}, the N-band extinction toward these two regions is about 3.5 to 15 mag. Therefore, even a relatively bright cluster star (i.e., 20-30 mJy at 10 $\micron$) will be undetectable in our observations due to this order of extinction.

\subsection{HD 259431}
HD 259431 is known to be closely associated with a nearby molecular cloud. In the CO survey centering on the Monoceros OB1 region by \citet{oliver1996}, HD 259431 appears to lie on the SE edge of a cloud. It also creates a reflection nebular (NGC 2247) that is noticeable in the POSSII/DSS2 optical images. In our view, these observations, as well as the youthfulness of the object, are suggestive of a stellar environment similar to what we find for the MWC 1080 system. Further, it seems that a bipolar cavity has been created by HD 259431 in its parent cloud via the outflow/wind, too. This speculation is in agreement with the diffuse mid-IR emission we observed. In particular, the bow-like structure described in Section \ref{sec:mid_ir_morphology} is likely tracing the SE half of the cavity. The outflow axis is moderately inclined with respect to the line of sight. Consequently, the NW half of the bipolar structure is obscured and invisible in our images. This inclination effect is able to account for the different mid-IR morphologies between HD 259431 and MWC 1080. To confirm the picture suggested here for HD 259431, however, we will need high-resolution CO maps (by SMA or ALMA) that allow a direct comparison with our deep mid-IR images. 

\section{Summary and conclusions}
We obtained deep images of MWC 1080 and HD 259431 at multiple wavelengths in the 10-$\micron$ and 20-$\micron$ bands. Our goal has been to shed new light on the nature of these two Herbig Be stars, in particular, the nature of their extended mid-IR morphologies, which may illuminate this key phase in the evolution of intermediate-mass star systems. Our data and analyses suggest that, in both systems, the observed diffuse emission is not confined to a simple disk or envelope, but originates in a more complex environment. We found that the filamentary, hourglass-shaped emission around MWC 1080 traces the internal surfaces of a gas cavity cleared by the bipolar outflow originating at or near the star MWC 1080A. The brightest part of the extended emission is elongated perpendicular to the hourglass axis, and can be easily confused with an edge-on disk or an envelope in low-resolution observations. In fact, the mid-IR emission appears to be defining in a striking way the geometry of the outflow cavity nearest the main star MWC 1080; in effect, the mid-IR emission is like a ``belt'' wrapped around the narrow waist at the base of the northern and southern outflows. The northern and southern edges of that belt of emission are diffuse and appear to trace very well the cavity walls in the vertical direction (i.e., along the flow) out to some distance from the star. Our work helps define the geometry of an active star formation region uncommonly well, and promotes future studies on the evolution of the MWC 1080 cluster influenced by the outflow and outflow-induced effects such as turbulence and gas clearing. Based on the similarities (morphology, connection to a molecular cloud, etc.) between HD 259431 and MWC 1080, we speculate that the filamentary emission found around HD 259431 is also originating in a cavity produced by outflow, although future high-resolution CO observations are needed to confirm this scenario. 

\acknowledgments
We thank the anonymous referee for the comments and suggestions that have greatly improved the clarity of this paper. This research has been supported in part by NSF grants AST-0908624 and AST-0903672 to C.M.T. This study is based on observations obtained at the Gemini Observatory through the program GN-2005B-Q-10 and GN-2010B-C-8. Gemini observatory is operated by the Association of Universities for Research in Astronomy, Inc., under a cooperative agreement with the NSF on behalf of the Gemini partnership: the National Science Foundation (United States), the Science and Technology Facilities Council (United Kingdom), the National Research Council (Canada), CONICYT (Chile), the Australian Research Council (Australia), Minist\'erio da Ci\.encia e Tecnologia (Brazil) and Ministerio de Ciencia, Tecnolog\'ia e Innovaci\'on Productiva (Argentina). 

Facilities: \facility{Gemini:North(Michelle)}

\clearpage

\begin{deluxetable}{l c c c c c c c l}
\tabletypesize{\scriptsize}
%\rotate
\tablecaption{Physical properties of program stars\label{tab:objects}}
\tablewidth{0pt}
\tablehead{
\colhead{Source} & \colhead{Spec.} & \colhead{$d$} & \colhead{log $T_{eff}$} & \colhead{log $L$} & \colhead{$M$} & \colhead{Age} & \colhead{$A_V$} & \colhead{Ref} \\
\colhead{ } & \colhead{Type} & \colhead{(pc)} & \colhead{(K)} & \colhead{($L_{\odot}$)} & \colhead{($M_{\odot}$)} & \colhead{(Myr)} & \colhead{ } & \colhead{ }
}
\startdata
MWC 1080 & B0e & 2.2k & 4.5 & 4.0 & 20.6 & $\sim$0.1-1.0 & 4.4-5.4 & 1-5\\
HD 259431 & B5e & 290 & 4.2 & 2.5 & 4.4 & $\le$1.0 & 0.88 & 6-9 \\
\enddata
\tablerefs{
(1) \citealt{cohen1979}; (2) \citealt{hillenbrand1992}; (3) \citealt{hillenbrand1995phd}; (4) \citealt{levreault1988}; (5) \citealt{abraham2000}; (6) \citealt{bouret2003}; (7) \citealt{brittain2007}; (8) \citealt{vandenancker1998}; (9) \citealt{bertout1999}.
}
\end{deluxetable}

\clearpage

\begin{deluxetable}{l l c c c c c c c}
\tabletypesize{\scriptsize}
%\rotate
\tablecaption{Log of observations\label{tab:obs_log}}
\tablewidth{0pt}
\tablehead{
\colhead{Program ID} & \colhead{UT date} & \colhead{Filter} & \colhead{$\lambda$} & \colhead{$\Delta\lambda$} & \colhead{Integration} & \colhead{PSF/flux} & \colhead{Air mass} & \colhead{PSF FWHM} \\
\colhead{ } & \colhead{ } & \colhead{ } & \colhead{($\micron$)} & \colhead{($\micron$)} & \colhead{(s)} & \colhead{standard} & \colhead{} & \colhead{(arcsec)}
}
\startdata
\multicolumn{9}{l}{\textbf{MWC 1080}} \\
GN-2005B-Q-10 & 2005 Aug 6 & Si-5 & 11.6 & 0.9 & 1192 & HD 5234 & 1.56 & 0.36 \\
                                     & 2005 Aug 6 & Qa & 18.5 & 0.9 & 626   & HD 3712 & 1.33 & 0.54 \\
                                     & 2005 Aug 6 & N'  & 11.2 & 2.1 & 706   & HD 5234 & 1.56 & 0.37 \\
                                     & 2005 Aug 16 & N'  & 11.2 & 2.1 & 376   & HD 5234 & 1.33 & 0.35 \\

\multicolumn{9}{l}{\textbf{HD 259431}} \\
GN-2005B-Q-10 & 2005 Dec 12 & Si-5  & 11.6 & 0.9 & 1192 & HD 49293 & 1.13 & 0.38 \\
                                     & 2005 Dec 28 & Qa  & 18.5 & 0.9 &  1200 & HD 49293 & 1.03 & 0.53 \\
                                     & 2005 Dec 28 & N'  & 11.2 & 2.1& 600 & HD 49293 & 1.38 & 0.38 \\
                                     & 2005 Dec 29 & N'  & 11.2 & 2.1& 600 & HD 49293 & 1.05 & 0.35 \\
GN-2010B-C-8   & 2011 Jan 29 & Si-4 & 10.3 & 1.0 & 900 & HD 49161 (PSF) & 1.28 & 0.33 \\
                                     &                        &           &       &          &         & HD 48433 (Flux) &           & \\
\enddata
\end{deluxetable}

\clearpage

\begin{deluxetable}{l c c c c c c}
\tabletypesize{\scriptsize}
%\rotate
\tablecaption{HD 259431 flux densities\label{tab:hd259431_photometry}}
\tablewidth{0pt}
\tablehead{
\colhead{Source} & \multicolumn{2}{c}{Offset\tablenotemark{a}} & \multicolumn{4}{c}{$F_{\nu}$} \\
\colhead{ } & \colhead{RA} & \colhead{DEC} & \colhead{10.3 $\micron$} & \colhead{11.2 $\micron$} & \colhead{11.6 $\micron$} & \colhead{18.5 $\micron$}
}
\startdata
\multicolumn{7}{c}{\textbf{Aperture radius = 1\farcs5}} \\
HD 259431 & 0 & 0 & 7.6$\pm$0.8 Jy & 6.6$\pm$0.7 Jy & 6.4$\pm$0.7 Jy & 5.5$\pm$0.6 Jy \\
 & & & & & & \\
\multicolumn{7}{c}{\textbf{Aperture radius = 2$\times$PSF FWHM}} \\
HD 259431 & 0 & 0 & 6.4$\pm$0.6 Jy & 5.9$\pm$0.6 Jy & 5.5$\pm$0.6 Jy & 5.2$\pm$0.5 Jy \\
Source B & -0.8 & 3.0 & 28$\pm$3 mJy & 26$\pm$3 mJy & 29$\pm$3 mJy & 13$\pm$5 mJy
\enddata
\tablenotetext{a}{Offsets in arcsec from HD 259431.}
\tablecomments{10\% photometric uncertainty assumed.}
\end{deluxetable}

\clearpage

\begin{deluxetable}{l c c c c c}
\tabletypesize{\scriptsize}
%\rotate
\tablecaption{MWC 1080 flux densities\label{tab:mwc1080_photometry}}
\tablewidth{0pt}
\tablehead{
\colhead{Source} & \multicolumn{2}{c}{Offset\tablenotemark{a}} & \multicolumn{3}{c}{$F_{\nu}$} \\
\colhead{ } & \colhead{RA} & \colhead{DEC} & \colhead{11.2 $\micron$} & \colhead{11.6 $\micron$} & \colhead{18.5 $\micron$}
}
\startdata
\multicolumn{6}{c}{\textbf{Aperture radius = 1\farcs5}} \\
MWC 1080 A+B & ... & ... & 16.6$\pm$1.7 & 16.9$\pm$1.7 & 13.8$\pm$1.4 \\
 & & & & & \\
\multicolumn{6}{c}{\textbf{Aperture radius = 2$\times$PSF FWHM}} \\
MWC 1080 A & 0 & 0 & 13.1$\pm$1.3 Jy & 13.3$\pm$1.3 Jy & 9.1$\pm$0.9 Jy \\
MWC 1080 B & -0.8 & -0.1 & 3.5$\pm$0.4 Jy & 3.6$\pm$0.4 Jy & 4.7$\pm$0.5 Jy \\
MWC 1080 E & 5.1 & 0.1 & 0.28$\pm$0.03 Jy & 0.31$\pm$0.03  Jy & 0.26$\pm$0.03  Jy \\
Source 1 & 6.8 & 1.2 & 58$\pm$7 mJy & 70$\pm$9 mJy & 67$\pm$11 mJy \\
Source 2 & 5.3 & -2.7 & 15$\pm$3 mJy & 20$\pm$4 mJy & $<$12 mJy \\
Source 3 & -1.6 & -4.7 & 13$\pm$2 mJy & 22$\pm$3 mJy & 23$\pm$6 mJy \\
Source 4 & -2.4 & -3.0 & 6$\pm$2 mJy & 11$\pm$2 mJy & $<$12 mJy \\
Source 5 & -6.1 & -5.1 & 4$\pm$2 mJy & 7$\pm$2 mJy & $<$12 mJy \\
Source 6 & -6.6 & 7.9 & 5$\pm$2 mJy & 6$\pm$2 mJy& $<$12 mJy \\ 
\enddata
\tablenotetext{a}{Offsets in arcsec from MWC 1080A.}
\tablecomments{10\% photometric uncertainty assumed.}
\end{deluxetable}

\clearpage
  
\clearpage

\clearpage

\clearpage

\clearpage

\clearpage

\end{document}